# The Immune System: look who's talking

Paolo Tieri
"L. Galvani" Interdept. Center, University of Bologna, Bologna, Italy
Dept. of Experimental and Diagnostic Medicine, University of Ferrara, Ferrara, Italy

**Abstract**

Human language and its governing rules present a number of analogies with the organization and structure of communication and information management in living organisms. This chapter will provide a short general introduction about grammar, as well as a brief explanation on how linguistic approaches effectively contaminate scientific practice, and, finally, how they can also provide systems biology with further tools and paradigms to analyse emergent behaviours and interactions among the components of a biological system.

## X.1 Talking, listening, acting

Essential life functions, as well as life itself, rely on intense and continuous collaborative effort by all the players in an organism. In complex systems, interdependencies among entities are so profound and substantial to require high communication skills, demanded at all levels. Information organization, storage, exchange and elaboration in living organisms, as well as in many other complex systems, seems to be organized in a coherent hierarchical scheme, from DNA to proteins, from cells to organs to the whole individual being. A similar arrangement of information can be found in the main way used by humans to communicate and interact: the spoken and written language (Editorial, 2002; Searls, 2003).

Language, the main system humans use to communicate, is – simplistically - composed of symbols and rules which govern their associations in such a way that allows them to carry information. According to several hypotheses that concern a fascinating discipline, the *linguistics,* these rules, generally referred to as grammar, seem to emerge spontaneously from inherent physical and mental characteristics of *Homo sapiens*. It is clear that these intriguing hypotheses are impossible to treat here due to their vastness and complexity. The architecture of human language presents a number of analogies with the organization and structure of communication and information management in living organisms. The following is a short description of how the various grammars of human languages formally represent these organizing principles, and how the introduction of approaches pertaining to linguistics contaminate and feed some fields of life science by providing effective paradigms for the analysis of emergent behaviours and interactions among the components of a biological system.

## X.2 Grammars, or hierarchical protocols to organize words into information

Human language naturally evolved for general-purpose communication among humans. Even though the exact definition is still under debate, nowadays linguists commonly refer to *natural language* as any spoken or written human language. Due to its typically complex characteristics, linguists have not yet fully understood all aspects of natural language, whose features seem to be deep-rooted in, and to depend on human vocal system as well as on human brain and human mind functionalities. Human languages such as for example Latin, English, Italian, Arabic and Japanese, as well as their regional varieties and dialects, are considered natural languages (Baker, 2003) (as distinguished from *artificial languages*, expressly constructed from individuals or groups for some particular use, such as Esperanto, for example).

The utilization of natural languages is essentially formalized in their grammars. Grammar, as mentioned, is the study of the rules of a language and it includes *morphology*, *syntax*, *semantics* and *pragmatics*. The focus of each of these subfields is briefly discussed below, which by no means should be considered exhaustive.

*Morphology* is the study of, and refers to, the *word structure.* The definition of "word" can be more complex than one can think. From a lexical point of view (different, for example, from *orthographical* or *phonological* acceptations), it is generally assumed that a word is the smallest unit of a lexicon or vocabulary and that it is able to carry a more or less recognizable function or meaning. All words or lexical items in a language can be considered as belonging to a dictionary of autonomous-existing, acceptable and currently spoken units, in turn consisting of chained characters taken from an *alphabet* or from a set of written symbols (*letters,* for example). It is also important to consider that words and their meanings can undergo changes imposed by the context and by their use.

As anyone can experience in everyday life, a coherent, logic and purposeful message or expression of any kind has a significance that cannot be inferred by the meaning of the individual words that make it up. *Syntax* refers exactly to the relations governing the way words combine with each other to form an understandable *sentence*, a string of words grammatically complete and potentially able to convey some meaning to the person receiving it (the word syntax comes from the Greek words συν -syn-, "together", and τάξις -táxis-, "arrangement", "placement").

*Semantics* pertains to the *meaning* carried by a set of juxtaposed words to form a sentence, or by another set of signs. The purpose of communication is to exchange meaningful messages in order to interact and coordinate with each other. Therefore, the importance of the meaning of a sentence is as clear as the saying "*you should keep on reading this chapter*", whose coherence and message is (hopefully...) caught by the person reading it. When compared with: "*a rope will ask you for some deeper clouds*", we can see that this last sentence is syntactically correct and formed by common and proper words but at the same time meaningless and illogic (at least in the common, everyday context, language and experience...).

Finally, *pragmatics* refers to the practical use of signs in different contexts. Its scope is to link the *literal meaning* of sentences to the *factual information* that the agent, through the communicative act, is trying to *convey*. As, once again, the common personal experience can demonstrate, the true meaning of a single sentence can be interpreted in different ways, depending on the context or on the person who receives it. It is easy to understand that the same sentence can carry diverging meanings based on different circumstances.

### *X.3 Grammatical structures to organize the biological matter?*

From the top level where organisms are classified into kingdoms, families and species, down to the molecular level of organic and bio-chemistry, there are several intermediate degrees in the organization of biological matter.

As an analogy, starting with the lowest levels and with the basic entities of life, one can see how the analogous of the *words* in biology can possibly be found in the *molecular domains*. Molecular domains are compact units with an independent stability, the autonomous folding modules of a protein. Like words of a language, molecular domains exist as basic functional units and they are, in turn, composed by chained amino acids that can be seen as the characters of our "life alphabet". When studying their *morphology*, in the same way we consider words, what it is important in molecular domain is their shape, their structure and pattern.

Moving on to the next level of organization, what has been said for linguistic syntax (protocols for assembling words into sentences) is also valid for the biological domain: the biological correspondent of a *sentence* at this level can actually be envisaged in the *protein* folded in the so-called *tertiary structure*. A protein is constituted by juxtaposed molecular domains (words) linked and stabilized by chemical bonds, bridges and interactions (conjunctions, punctuations...) that form a defined three-dimensional structure strictly related to its biological function (the sentence and its meaning).

As cells talk to each other by means of exchanged proteins, and proteins carry a message related to a specific function, it is arguable that cells obey to their semantics, intended as the "environment" where the relation between the message carrier and the function to be performed by the receiver are formalized. Since in any language a "mechanical" junction of words is not always related to a meaningful content, in the same way the true "meaning" of a protein can be linked to the *function* performed by the protein in relation to its final structure, its thermodynamic properties, interaction capabilities and potentiality to carry an understandable message from a cell to another (or among cellular compartments). Based on this capability, cells, as well as intracellular components, are able to set up a coherent conversation and perform a complex and coordinated response.

The problem with the *interpretation* of the message by the receiver and the consequent action (pragmatics) also exists in biology. There are, in fact, examples of different functions stimulated by the same protein when bound to different targets. Two relevant cases are those of the proteins NGF (nerve growth factor) and TNF-$\alpha$ (tumor necrosis factor alpha). NGF is able to bind two different receptors and thus to initiate different responses (Hempstead, 2006). Similarly, TNF-$\alpha$ can bind two different TNF receptors, p55 and p75, and initiate both gene expression, by activation of NFkB transcription factor, or apoptosis, by recruitment of TRADD complex and activation of upstream caspases (Abbas *et al.*, 2007).
Moreover, a wider interpretation of this analogy can take into consideration the ability to sense as "natural" or otherwise "non natural" molecules that are, or are not, in the proper compartment or context, as proposed in the well-known immunological hypothesis called "danger model" (Matzinger 1994, 1998), which suggests that immune system discerns in this way between what is dangerous or not dangerous to the host. For example, uric acid is endogenous and constitutively present in cells and its concentration increases when cells are injured. When released from dying cells, it appears to be misplaced and to stimulates dendritic cells, the receivers of the danger message, to mature and augments CD8+ T-cell responses to cross-presented antigens (Shi *et al.*, 2003).

The outline in Table 1 tries to summarize in a simple way the players in the two fields of linguistics and molecular biology. The blue rows represent the tentative common organizing principles that allow to scale the complexity among hierarchical levels and to give rise to the upper organized entity , which in turn becomes the basic unit for further scaling.

| **human being** (speaker/listener *agent*) | **cell** (secreting/targeted *agent*) |
|---|---|
| *Pragmatics: original meaning in the intention of the agent, different interpretation depending on context* | |
| conversation, discourse | cascade, pathway, network |
| *Semantics: message exchange between entities; the things, actions, objective situations (actual or potential) related to the information received* | |
| phrase, sentence | protein |
| *Syntax: arrangement of recognizable patterns* | |
| word | protein domain |
| *Morphology: basic unit structure, lexicon of valid units (words or stable amino acid configurations)* | |
| character, alphabet | base triplets, amino acid |

*Table 1: agents, organizing and analysing principles, and relative hierarchy; complexity increasing from bottom to top.*

## X.4 From debatable analogies to some useful science

As suggested by SEARLS and collaborators (Dong and Searls, 1994; Rawlings and Searls, 1997; Searls 1997, 2001, 2002, 2003), today that human genome can be considered unveiled in its sequence, it is clearer that the analogy with the literature overcomes the simple symbolic meaning used since the beginning of molecular biology, when terms like translation or transcription were adopted to indicate the various processes of protein biosynthesis from DNA. As a matter of fact, several problems have interestingly been faced in molecular biology using approaches and tools borrowed from linguistics, therefore demonstrating that the analogies shown here have overcome the mere conceptual field and have practically and successfully contaminated the scientific practice.

As GIMONA asserts (Gimona, 2006), "*a key theoretical principle for understanding an unknown language is the recognition of syntactic patterns. For proteins, these patterns might be similarities in sequence, or structure, or both*". It is exactly following this approach and exploiting such analogies that, in their very recent work on the design of antimicrobial peptides (AmPs, small natural proteins that immune system employs to fight bacterial infections), LOOSE and collaborators (Loose *et al.*, 2006) started treating AmPs as pieces of a conversation written in a out-and-out biological language. They syntactically analysed their amino acid sequence and their structure to recover a set of grammar rules able to produce a general description of this language. In this way and with a grade of success, the authors were able to identify the construction principles of the existing peptides and, by attaining to the same grammar, to create new synthetic peptides that showed a certain antimicrobial effectiveness. After uncovering the syntax, it has simply been possible to write new coherent and meaningful sentences in the form of unnatural, synthetic and effective AmPs.

With quite a similar strategy, PRZYTYCKA, AURORA, SRINIVASAN and ROSE (Przytycka *et al.*, 1999, 2002) have explored the possibility that protein folding processes are governed by a limited set of rules able to provide a wide variety of domain structures as all-β folds (a class of protein structural domains in which the secondary structure is entirely composed of β-sheets, such as the one in the immunoglobulin fold). The researchers argue that by applying just four simple constructing principles derived from repetitively observed patterned structures, all-β folds could be reconstructed starting with the simpler basic units in which they were decomposed. By observing how the rules correctly describe the folding of all-β folds with the known hierarchical organization of protein domains, they concluded that the existence of a grammar for protein composition has implications in folding and domain evolution.

In his interesting view, Jı (Ji, 1997) further exploits the analogy between linguistics and biology. He proposes an isomorphism (a closer, one-to-one correspondence between the elements of two sets) amongst cell language and human language, thus suggesting -with his interesting discussion- the existence of two classes of nucleotide sequences: one of structural genes for the *lexicon* (classical coding regions), and one of "spatiotemporal" genes coding for the *grammar* of cell language (classical non-coding regions). Interestingly, the hypothesis of this complete cell language theory brings us to the prediction of this "second genetic code" that should map the nucleotide sequences onto folding patterns of DNA and control the dynamics of gene expression.

Cell-to-cell language in bacterial populations impinges upon and regulates what appears to be a high-level coordination. *Quorum sensing* is the ability of bacteria to communicate and coordinate social behaviour and specific development via signalling molecules. It is widely accepted that specific proteins, sentences understandable to cells, act in this instance as a universal lexicon that mediates intra- and inter-species bacterial behaviour (Waters and Bassler, 2005). There are many examples of regulation molecules that coordinate the social and pluricellular behaviour of bacteria, almost like a *motto* under which the cohorts of individuals organize themselves to form sophisticated colonies with a specific well-defined behaviour. Communication in their particular language, here in the form of quorum sensing systems, is the mechanism that allows bacteria to switch functions from individual organism to multicellular organisms and, in this way, to further scale in complexity and evolution.

Such promising results and examples unquestionably exert over researchers an exciting appeal about the potential use of much older, well proofed and refined linguistic techniques and approaches. In fact, it is worth to mention that the formulation of the rules of Sanskrit morphology by the Indian grammarian Pāṇini, one of the earliest linguistic works, seems to go back straddling fifth and sixth century BC. During the past several decades this interest has been corroborated by the continuously increased attention, under many different perspectives, read in the major scientific and specialised headings (Brendel and Busse 1984; Botstein and Cherry, 1997; Boguski, 1999; Benner and Gaucher, 2001; Koonin *et al.*, 2002; Banavar *et al.*, 2004, just to cite a few).

### *X.5 Towards integration and dynamic analysis of signalling networks*

It is well known that complex behaviour emerges from interaction of many simpler agents that self-organise in multiple hierarchical scales: in every single level it is possible to identify the basic entity that, by combining, communicating and interacting with other similar entities gives rise to the higher, next-level entity, which in turn is concerted with other same-class entities. This process is repeated with increasing complexity of the organism or system and possibly of its tasks.

The Immune System (IS) is a mixture of many types of proteins, cells, tissues, and organs which strenuously and frenetically interact in an elaborate and dynamic network. Cohen and collaborators (Cohen 1992, Cohen *et al.* 2004, Cohen 2006, Cohen *et al.* 2007) have deeply investigated with their seminal work the cognitive features of the IS. Focusing on its marked information-exchange capabilities, it is easy to realise how its rapid and coordinated tasks and responses require a high level of communication. Soluble mediators such as cytokines, chemokines and hormones implement cellular communication among a numerous set of different cell types by patrolling, or resting, or acting upon a variety of distinct activation states, both at short range (the so-called *autocrine* action, or self stimulation, and *paracrine* action, or adjacent cell stimulation) and long range (immune, endocrine and nervous systems interactions). These mediators have a fundamental role in regulating the reaction of the IS: its function strongly rely on the intense intercellular crosstalk (Abbas *et al.*, 2007; Janeway *et al.*, 2004) as well as, probably, on the interference and intrinsic noise of the system (Edelman and Gally, 2001; Tieri *et al.*, 2007).

Communication among the IS components seems to be based, as in any other form of communication, on precise rules: similar to sentences in a conversation, the exchange of signals and proteins must be comprehended by cells within definite limits of interpretation (binding affinity) that are not beyond their reading capability, in order to prevent confusion, disorder and malfunctions (many diseases are commonly said to be generated and derived from a chemical message that is erroneously interpreted due to cell's incapacity, or because the message is corrupted and misleading, or arrived in the wrong time or quantity).

At this point, it appears evident that in observing the phenomenon of cell-to-cell communication we find ourselves in front of an out-and-out conversation that carries a definite message, undertaken in a language we are just starting to translate and comprehend, which is targeted to the precise scope of acting for the sake of definite behaviour coordination. As common logic suggests, this kind of action requires the existence of spatial and temporal limitations that give coherence and sense to the speech, enforced in a manner dictated by biochemical and biological constraints (binding affinity, decay, function loss, change of state, context, etc.).

In studying the structure in itself of such biological conversation, one should be concerned not only about the carried message and its intrinsic quality, but also about quantities involved in the molecular crosstalk. As a matter of fact, a signalling pathway is often initiated only when cell surface receptors have bound, in a restricted time windows, a definite number of complementary ligand molecules, thus generating an "impulse" that exceeds a definite intensity, a threshold necessary for the signal to be comprehensible and intentional, almost like a biological *repetita iuvant*. Based on this perspective and on our intent, our work on cell-to-cell signalling analysis (Tieri *et al.*, 2005) is addressed to study the communication channels among cells, considering not only the single signal molecule and the message it is carrying but also the different channels that cells can use and exploit to convey the information among the system's components.

Besides, as system's complexity increases, one can easily realize how information and language become more and more important in order to coordinate and manage the entire situation. In this direction, many interdisciplinary efforts appear to be lavished in linguistics and its subfields. Besides, advanced statistical approaches are not new in linguistics: among the topics discussed in the most recent conferences in the discipline, it is worth to mention the application of Kolmogorov's complexity theory to disentangle the diverse levels of complexity in the different levels of language processing (see for example the COLING conference series). Hopefully, these approaches are now cross-fertilizing the fields of molecular and systems biology; successful analysing methodologies are expected to bounce back and forth between biologists and linguists to give rise to powerful research tools and paradigms in both disciplines. At heart, we humans talk to each other thanks to our cells talking to each other.

*Acknowledgements*

I wish to thank Jonathan Timmis, Stefano Salvioli, Gastone Castellani, Claudio Franceschi and Joseph and Loredana Iannuzzo for useful discussion, suggestions and comments.